\documentclass[12pt]{iopart}
\pdfoutput=1

\usepackage{iopams}  
\usepackage{graphicx}
\usepackage{caption}
\usepackage{subfig}
\usepackage{color}
\usepackage[margin=1in]{geometry}
\usepackage{multirow}
\usepackage{tabularx}

\begin{document}

\title[Time-dependent simulation]{Time-dependent simulation of particle and displacement currents in THz graphene transistors}

\author{Z. Zhan, E. Colom\'es,  A. Benali, D. Marian  and X. Oriols}
\address{Depertament d'Enginyeria Electr\`onica, Universitat Aut\`onoma de Barcelona, Bellaterra, Spain}
\ead{xavier.oriols@uab.cat}
\vspace{10pt}
\begin{indented}
\item[]December 2015
\end{indented}

\begin{abstract}
Although time-independent models provide very useful dynamical information with a reduced computational burden, going beyond the quasi-static approximation provides enriched information when dealing with TeraHertz (THz) frequencies. In this work, the THz noise of dual-gate graphene transistors with DC polarization is analyzed from a careful simulation of the time-dependent particle and displacement currents. From such currents, the power spectral density (PSD) of the total current fluctuations are computed at the source, drain and gate contacts. The role of the lateral dimensions of the transistors, the Klein tunneling and the positive-negative energy injection on the PSD are analyzed carefully. Through the comparison of the PSD with and without Band-to-Band tunneling and graphene injection, it is shown that the unavoidable Klein tunneling and positive-negative energy injection in graphene structures imply an increment of noise without similar increment on the current, degrading the (either low or high frequency) signal-to-noise ratio. Finally, it is shown that the shorter the vertical height (in comparison with the length of the active region in the transport direction), the larger the maximum frequency of the PSD. As a byproduct of this result, an alternative strategy (without length scaling) to optimize the intrinsic cut-off frequency of graphene transistors is envisioned.

\end{abstract}

\maketitle

% % Uncomment for PACS numbers
%\maketitle

%%%%%%%%%%%%%%%%%%%%%%%%%%%%%%%%%%%%%%%%%%%%%%%%%%%%%%%
%%%%%%%%%%%%%%%%%%%%%%%%%%%%%%%%%%%%%%%%%%%%%%%%%%%%%%%
%%%%%%%%%%%%%%%%%%%%%%%%%%%%%%%%%%%%%%%%%%%%%%%%%%%%%%%
\newpage

\tableofcontents

\section{Introduction}

Although many new materials have been proposed as candidates to substitute the old-fashioned Silicon field effect transistors (FETs), a recent article concluded that : \emph{``many such saviours have come and gone, yet the reliable silicon CMOS continues to be scaled and to reach even higher performance levels ''} \cite{Ferry2008}. Among these new materials, graphene is expected to have a great potential impact in our society, in general, and in electronics, in particular \cite{Novoselov2009, Schwierz2010}. Graphene is a single 2D layer of carbon atoms with a hexagonal lattice \cite{Novoselov2009}. It has a linear energy-momentum dispersion (which provides massless Dirac fermions), an extraordinary elasticity (allowing flexible electronics) and extremely large electrical conductivity (with electron velocities of $10^{6}\;m/s$) \cite{Geim}. However, graphene has a zero bandgap implying a small on-off ratio for graphene digital FETs. In addition, reliable techniques to create a sizeable gap degrade a lot the properties mentioned above. Therefore, it seems that successful graphene logic applications are not currently feasible. On the contrary, the large conductivity of graphene is very welcome for (small-signal) radio frequency applications (such as amplifiers or mixers) \cite{Schwierz2010, Schwierz2011, Sordan} which are not required to switch off and can benefit from the high mobilities offered by graphene. 

Although the best performance of nowaday radio-frequency graphene transistors is still quite below the one obtained from Silicon and III-V HEMTs \cite{Schwierz2010},  significant progress has been made since the experimental demonstration of the first GigaHertz graphene transistors in 2008 \cite{Meric}. Most notably, a research group reported graphene FETs breaking the 100-GHz cut-off frequency ($f_{T}$) mark in 2010 \cite{Lin}. Furthermore, only a few months later, researchers demonstrated \cite{Liao} a graphene FET that has a $f_{T}$ of 300 GHz. On the theoretical side, a very simple estimation of the intrinsic cut-off frequency as the inverse of electron transit time shows that one can easily reach frequencies higher than 1 TeraHertz (THz) (active region shorter than $10^{-6}\;m$ with electron velocity on the order of $10^{6}\;m/s$). \emph{Are such simple intrinsic high-frequency predictions really achievable? What noise is expected at such frequencies?} Usually, predictions of the high-frequency behavior of graphene devices are studied through its simulations under quasi-static approximations, where the time derivative of the electric and magnetic fields is neglected.  There are many successful examples in the literature on how quasi-static approaches are still capable of getting THz information of electron devices. Among many others, we mentioned those based on time-independent solutions of the Non-Equilibrium Green's function framework owing to the Klimeck's group with the NEMO simulator \cite{Kim} or to Fiori and Iannaccone's group with the NANOTCAD ViDES simulator \cite{Logoteta}. In this regard, the \emph{ab initio} (time-independent ground-state) density functional theory (DFT) has also been successfully used in the literatures for such graphene THz predictions \cite{Zheng, Chanana}. The strategy of all these steady-state simulators, for example, for predicting the cut-off frequency is, first, simulating the values of currents and charges from their powerful quantum steady-state simulators, then, calculating transconductances and capacitances from the currents and charges dependence on the (gate) voltage, respectively, and finally, plugging these last calculations of transconductances and capacitances into an analytical expression of the cut-off frequency (usually obtained from a small-signal circuit model). In any case, the procedure of getting AC properties from time-independent (steady-state) simulations has been demonstrated to be very successful, providing very-valuable physical insight of the high frequency problems while greatly reducing the computational burden associated to explicit time-dependent simulations \cite{Fiori}. For example, these quasi-static approaches have clearly demonstrated that one of the most relevant drawbacks today for getting the expected theoretical THz  cut-off frequencies in graphene transistors is the (extrinsic) resistance between the metallic 3D contact and the 2D graphene channel \cite{Xia, Fiori}.  
 
This paper is aimed to carefully study the intrinsic THz performance of graphene transistors beyond the quasi-static approximation, by explicitly simulating the time-dependent displacement and particle  currents in the active region of graphene transistors. In our previous work, we have presented an original strategy to optimize radio frequency performance of gate-all around (GAA) quantum-wire Silicon FETs by modifying their lateral areas, without length scaling or mobility improvement \cite{Benali2013}. We have seen that, in the definition of the duration of the total current peak due to an electron travelling along the device, there are scenarios where the exact transit time of the electron is not at all a relevant parameter for $f_{T}$ estimations. We have proved \cite{Benali2013} that, for GAA quantum-wire Silicon FETs, the ultimate responsible of the high-frequency noise is not the electron transit time $\tau_e$, but a different time related to the duration of the total current peak $\tau_i$, while the electron is crossing the device. In this paper, we will show that similar arguments can be also applied to the intrinsic THz performance of graphene transistors. Along this work we refer to intrinsic modeling in order to emphasize that spurious effects (like the important drawbacks occasioned by the contact resistance mentioned above) are not considered in this work. We only deal with electron dynamics inside the device active region. The intrinsic Klein (Band-to-Band) tunneling and positive-negative energy injection on graphene transistors are carefully analyzed and we provide predictions on their effect on the graphene high-frequency performance. From the numerical simulations, we show that the unavoidable Klein tunneling and positive-negative energy injection in graphene structures increase the THz noise one order of magnitude without similar increment on the current, degrading the signal-to-noise ratio, in comparison with the simulation without Band-to-Band tunneling and only positive energy injection.
 
After this introduction, in section 2 and section 3, we provide some preliminary discussions about our simulation tool beyond the quasi-static approximation, i.e. through the explicit simulation of time-dependent particle and displacement currents. We also discuss the relation between electron dynamics and high-frequency spectrum depending on the fact of dealing with 2-terminal (resistor) or 3-terminal (transistor) devices. In section 4 and section 5, the BITLLES simulator \cite{bitlles} is used to study the intrinsic high-frequency performance of graphene FETs, in terms of semi-classical trajectories compatible with the energy-dispersion of the Dirac equation, Klein tunneling and positive-negative (kinetic) energy injection model. 

%%%%%%%%%%%%%%%%%%%%%%%%%%%%%%%%%%%%%%%%%%%%%%%%%%%%%%%
%%%%%%%%%%%%%%%%%%%%%%%%%%%%%%%%%%%%%%%%%%%%%%%%%%%%%%%
%%%%%%%%%%%%%%%%%%%%%%%%%%%%%%%%%%%%%%%%%%%%%%%%%%%%%%%

\section{Preliminary discussions}

Since most of the predictions of high-frequency performance of graphene transistors have been obtained from the quasi-static approximation with time-independent models, some preliminary discussions about the formalism and what kind of new results can be accessible with this time-dependent (dynamic) simulation of particle plus displacement currents are mandatory. 

The motivation for focusing our attention on the displacement current is the following. The particle current (i.e. the flux of electrons on a particular surface) is basically related to the injection rate of electrons from the contact and its order of magnitude is not modified when the input signal frequency of the device is increased. On the contrary, the displacement current (i.e. the time-derivative of the electric field on a particular surface) is roughly proportional to the input signal frequency. Therefore,  at very low frequencies only the particle current is relevant, while the displacement current becomes negligible in front of the particle current. This is the typical working region of electronics. On the contrary,  at frequencies high enough, the displacement current (proportional to the frequency growth) becomes the only relevant in front of the particle current. This is the working scenario for electromagnetic applications. Certainly, in the frontier between typical electronics and electromagnetism there is a region where both displacement and particle currents become relevant. 

\subsection{The measured total current in different parts}

\begin{figure}[h!!!]
\centering
\includegraphics[width=0.85\columnwidth]{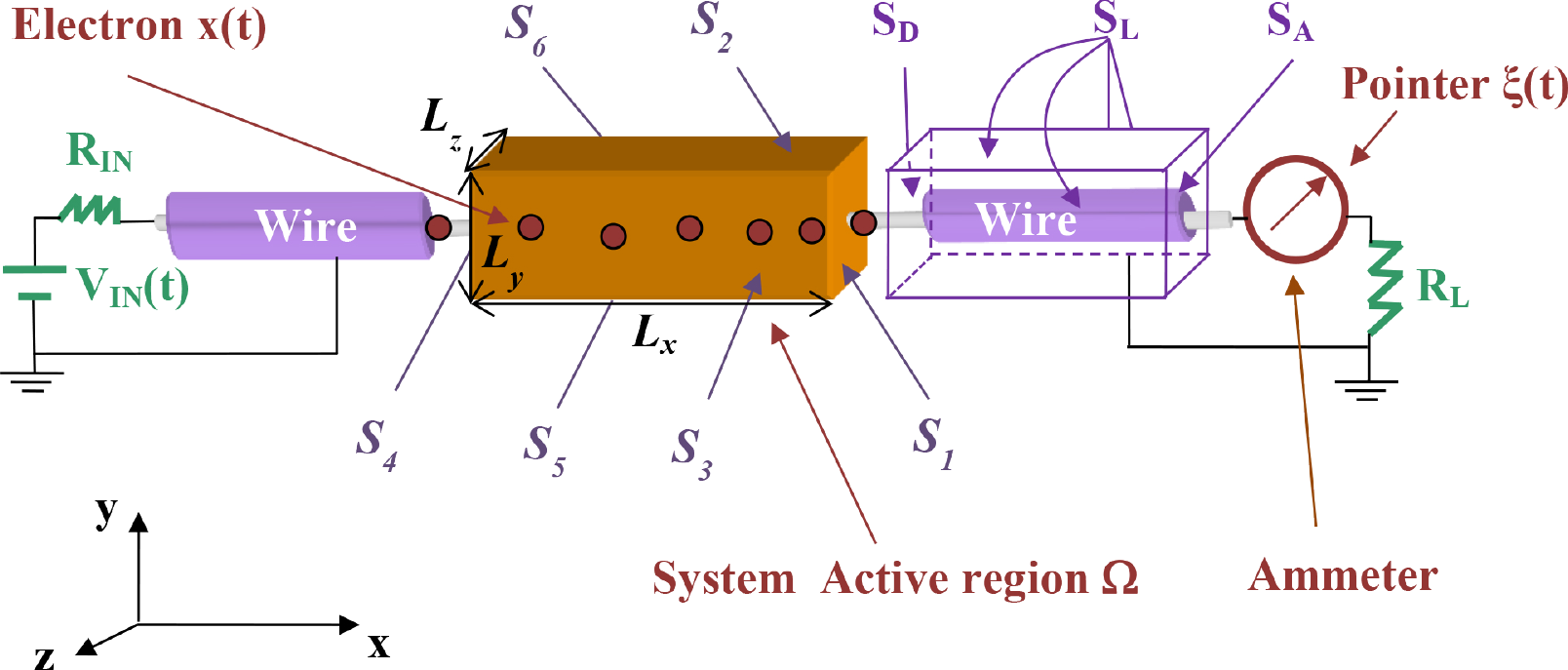}
\caption{Schematic representation of a typical electrical circuit used for studying the difference between the computed and the measured currents in electrical devices.}
\label{measurement}
\end{figure}
Let us discuss first the role of the displacement current on the measured value of the current that we get in a laboratory. It is common to compute the electrical current on the (simulated) surface $S_D$ ($S_D$=$S_1$ at the device contact) of the active region in \fref{measurement}, while a real measurement is performed on the (non-simulated) surface $S_A$ in the ammeter. \emph{Then, is the current on $S_D$ equal to that on the ammeter?}  In fact, these currents will only be identical if we consider the total current $I_T(t)=I_p(t)+I_d(t)$, where $I_p(t)$ and $I_d(t)$ are respectively the particle and displacement components. Owing to the current conservation law, an integral of the total current density $\vec{J_T}(\vec{r},t)$ on a closed surface, for example, in the wire $S=\{ S_D, S_A, S_L \}$ in \fref{measurement} is zero. We have defined $S_L$ as the surface parallel to the transport direction in the cable. In particular, for a cable we can assume $\int_{S_L}\vec{J_T}(\vec{r},t)d\vec{s}=0 $, so finally we get $\int_{S_D}\vec{J_T}(\vec{r},t)d\vec{s}=-\int_{S_A}\vec{J_T}(\vec{r},t)d\vec{s}$. The important point is that we have to simulate the total current (not only the particle current) on $S_D$ if we want to ensure that the simulated result is equal to the measured one on $S_A$. 

\subsection{The displacement and particle currents}

The standard expression of the displacement current, $I_d(t)$, evaluated on a surface $S_j$ of the active region $\Omega$ in \fref{measurement} is proportional to the time-derivative of the electric field:

\begin{equation}
I_{d}(t)  =  \int_{S_j}  \epsilon(\vec{r})\frac{d\vec{E}(\vec{r},t)}{dt} \cdot d\vec{s}, 
\label{displacement}
\end{equation}
where $\epsilon(\vec{r})$ is the inhomogeneous electric permittivity, and $\vec{E}(\vec{r},t)$ is the electric field vector at the position $\vec{r}$ at the time $t$. The particle current expression, $I_p(t)$, computed on $S_j$ is

\begin{equation}
I_{p}(t)  =  \int_{S_j}  \vec{J_c}(\vec{r},t) \cdot d\vec{s}=\lim_{\Delta t \to 0 } \displaystyle\sum_{m=1}^{N_p} \frac{q}{\Delta t} sign(\vec{v}_m),
\label{particle}
\end{equation}
where $\vec{J_c}(\vec{r},t)$ is the current density vector at the position $\vec{r}$ at the time $t$, the sum $N_p$ is the number of electrons that have crossed the surface $S_j$ during time step of the simualtion $\Delta t$, and $q$ is the electron charge without sign. The function $sign(\vec{v}_m)$ is equal to 1 when one electron leaves the volume $\Omega$ through the surface $S_j$, while $sign(\vec{v}_m)=-1$ when the electron enters.

\subsection{The Ramo-Shockley-Pellegrini theorem}

In order to discuss in detail the relation between the total (particle plus displacement) current and the dynamics of electrons inside a semiconductor device, using Green's second identity to create another electrokinematics equations, we provide an alternative expression (named Ramo-Shockley-Pellegrini (RSP) theorem \cite{Pellegrini}) for computing the time-dependent total current different from direct definitions of displacement current (equation \eref{displacement}) and particle current (equation \eref{particle}). We consider the parallelepiped of volume $\Omega=L_x \cdot L_y \cdot L_z$ limited by the closed surface $S$ which is composed of six rectangular surfaces $S=\{ S_1, S_2, \dots, S_6  \}$ as seen in the device active region of  \fref{measurement}.

The total time-dependent current in a surface $S_j$ is defined by the RSP theorem as $I_T(t)=\Gamma_j^q(t)+\Gamma_j^e(t)$, and with: 
\begin{eqnarray}
\fl \qquad \Gamma_j^q(t)=-\int_\Omega \vec{F}_j(\vec{r}) \cdot \vec{J}_c(\vec{r}, t) \; \mathrm{d}\mathit{v}=-\displaystyle\sum_{m=1}^{N_p} sign(\vec{v}_m) \; q \; \vec{F}_j(\vec{r}_m)\cdot \vec{v}_m ( \vec{r}_m),  \label{volume}\\
\fl \qquad \Gamma_j^e(t)=\int_{S_j} \epsilon(\vec{r}) \; \frac{\mathrm{d}V(\vec{r},t)}{\mathrm{d}t}\; \vec{F}_j(\vec{r}) \cdot  \mathrm{d}\vec{s},  \label{surface}
\end{eqnarray}
where $\vec{v}_m ( \vec{r}_m )$ is the $m$-th electron velocity, $V(\vec{r},t)$ is the scalar potential at position $\vec{r}$ and time $t$. The vector function $\vec{F}_j(\vec{r})$ is defined through an expression $\vec{F}_j(\vec{r})=- \mathbf{\nabla}\phi_j(\vec{r})$, where $\phi_j(\vec{r})$ is its scalar potential. Let us note that the terms $\Gamma_j^q(t)$ and $\Gamma_j^e(t)$ cannot be interpreted as particle current and displacement current, respectively. In fact, the term $\Gamma_j^q(t)$ includes itself the particle current and part of the displacement current altogether. For example, when an electron is not crossing a surface, say $S_j$, $\Gamma_j^q(t)\neq 0$ while $I_{p}(t) =0$.

\subsection{Is the electron transit time $\tau_e$ equal to the current peak time $\tau_i$ ?}

\begin{figure}[h!!!]
\centering
\includegraphics[width=0.80\columnwidth]{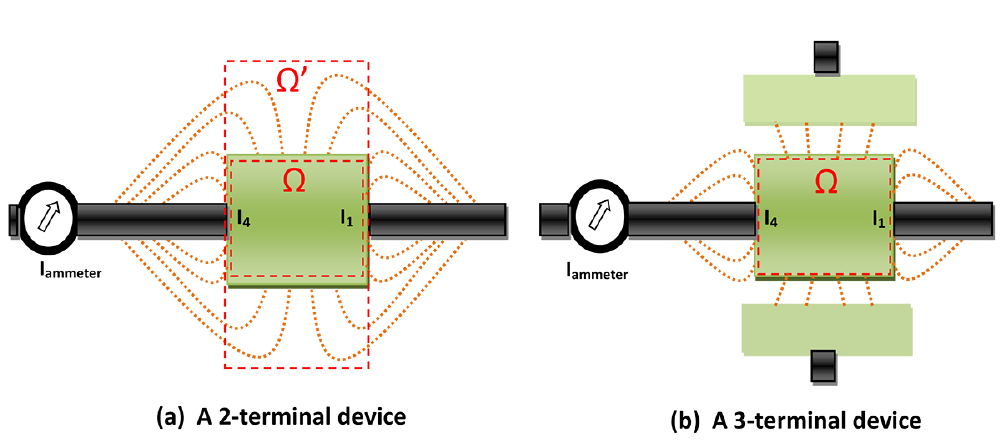}
\caption{(a) A 2-terminal device, we use the very large simulation box $\Omega'$ to compute the total current, i.e. $I_4$ on surface $S_4$, which is equal to the current $I_{ammeter}$ in the ammeter. (b) A 3-terminal device. The dashed lines represent the variational electric field lines.}
\label{figure2}
\end{figure}

The RSP theorem provides a very simple mathematical framework to answer this question. First, we focus on a 2-terminal device. The first version of the RSP theorem was presented by the work of Shockley \cite {Shockley} in 1938 and of Ramo \cite {Ramo} in 1939. They, separately, provided a simple expression for the  computation of the total (particle plus displacement) current flowing through a vacuum tube, i.e. the typical electron devices at that time. A vacuum tube can be roughly modeled as two infinite metallic plates separated by air. According to the \fref{figure2}(a), we name $S_4$ the left plate and $S_1$ the right plate, assume that an electron is moving inside volume $\Omega$. It can be demonstrated that we get $\vec F_4(\vec r) \approx-(1/L_x)\cdot\vec x$ when considering the volume $\Omega\rq{}$ (the bottom and upper surfaces are much smaller than the lateral ones) which captures all the lines of the electric field generated in the active region by the moving electron \cite{Benali2012}, where $L_x$  is the distance between plates and $\vec x$  is the unit vector in the transport direction perpendicular to the plates. Then, using equations \eref{volume} and \eref{surface} for just one electron moving with velocity $\vec v=\{v_x,0,0\}$ in the transport direction, the total current on $S_4$ during $0<t<\tau_e$ can be written as:
\begin {equation}
I_4(t) \approx \Gamma^{q}_4(t) \approx -q \frac {v_x(t)} {L_x},
\label{equation_32}
\end {equation}
being $q$ the (unsigned) electron charge. The current value is a constant while the electron is inside the $\Omega$ and its time-integral during $\tau_e=L_x/v_x$ gives the expected transmitted charge $-q$.  In addition, the currents\footnote[1]{ Except when specified the contrary, current in the text means total (displacement plus particle) current} on $S_1$ and $S_4$ are equal at any time. In this particular case, the relevant time for the peak current $\tau_i$ is roughly equal to the electron transit time $\tau_e=L_x/v_x$. \emph{Can we envision other scenarios where transient current collected on a particular surface  is not related to the electron transit time?} Below, we answer positively to this question. 

To go beyond the previous Ramo-Shockley result is mandatory to deal with, at least, a 3-terminal device that ensures the instantaneous current $I_S(t)$ in the source $S$ is not equal to that $I_D(t)$ in the drain $D$, while still satisfying the instantaneous current conservation.  For this reason, we consider the 3-terminal dual-gate FET, see, for example, volume $\Omega$ in \fref{figure2}(b), where the top and bottom surfaces are no longer smaller than the lateral surfaces. For that geometry of the volume $\Omega$, we get the expression $\vec F_4(\vec r) \approx -\alpha_x \cdot exp\;(\alpha_x(x-L_x))\cdot\vec x$ \cite{Benali2012}, where $\alpha_x=\sqrt{(\frac{1}{L_y})^2+(\frac{1}{L_z})^2}$ (being $L_y$ the vertical height and $L_z$ the width indicated in \fref{measurement}). $\vec F_4(\vec r)$ is not constant neither in modulus nor in direction. We can write the current on $S_4$ due to one electron moving in x direction with velocity $v_x$ directly from equations \eref{volume} and \eref{surface} as, 
\begin{equation}
I_4(t) \approx \Gamma^{q}_4(t) \approx -q\; v_x \; \alpha_x \; e^{v_x\;\alpha_x(t-\tau_e)}.
\end{equation}
We have seen clearly that the geometry of $\Omega$ has a clear influence on $\vec F_4(\vec r)$,  which, in turn, affects the current $I_4(t)$. Here, the electron transit time is different from the current peak duration, $\tau_e>\tau_i$. In the next section, we will analyze how the lateral surface of transistors (that do not affect the electron transit time $\tau_e$ in the transport direction) effectively affects the shape of the current pulse $\tau_i$.

%%%%%%%%%%%%%%%%%%%%%%%%%%%%%%%%%%%%%%%%%%%%%%%%%%%%%%%%
%%%%%%%%%%%%%%%%%%%%%%%%%%%%%%%%%%%%%%%%%%%%%%%%%%%%%%%%%
%%%%%%%%%%%%%%%%%%%%%%%%%%%%%%%%%%%%%%%%%%%%%%%%%%%%%%%%%
\section{Time-dependent simulation of the total current in graphene structures}

\begin{figure}
\centering
\includegraphics[width=0.65\columnwidth]{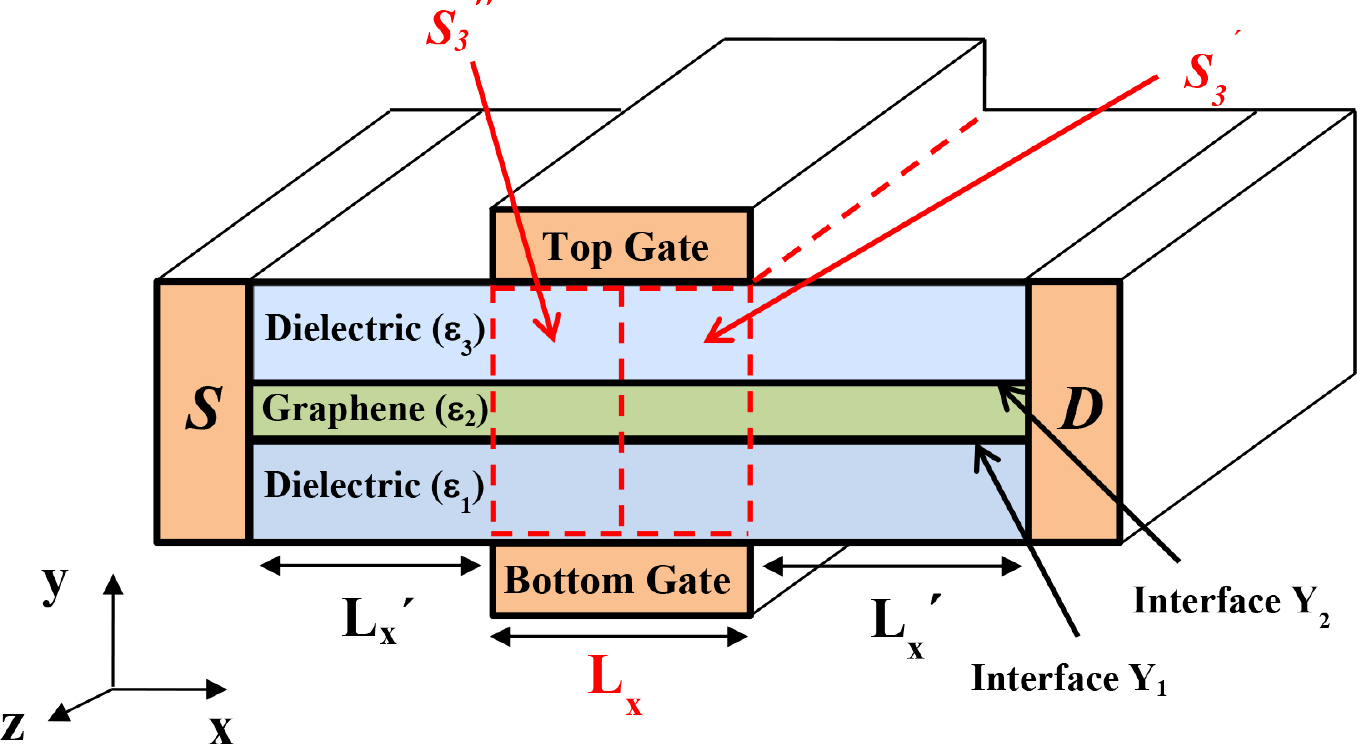}
\caption{Schematic representation of a graphene device (graphene channel is sandwitched between two dielectrics) in BITLLES simulator, whose volume $\Omega$ (with 3D structure plotted by dashed lines) is limited by a closed surface $S=\{ S_1, S_2, \dots, S_6  \}$ that is indicated in \fref{measurement}. In volume $\Omega$, the channel length $L_x=20\;nm$, the surface $S_3$ (and $S_6$) are divided into $S_3'$ ($S_6'$) and $S_3''$ ($S_6''$). Length $L_x'=30\;nm $.}
\label{figure3}
\end{figure}

After the general discussion presented in the previous section, we now provide some simple results focus on graphene transistors. We consider an active region of dual-gate graphene FETs plotted in \fref{figure3}. A large-area 2D graphene is placed as a channel, sandwiched between a top-gate dielectric and a bottom-gate dielectric. The active region is defined as the volume $\Omega$, which is limited by a closed surface $S=\{ S_1, S_2, \dots, S_6  \}$ with six surfaces defined the same as that in \fref{measurement}. 

\subsection{The role of different dielectric constants on the high-frequency behavior}

\begin{figure}
     \centering
     \subfloat[][$\varepsilon_2 \gg \varepsilon_1, \varepsilon_3 $]{\includegraphics[width=0.40\columnwidth]{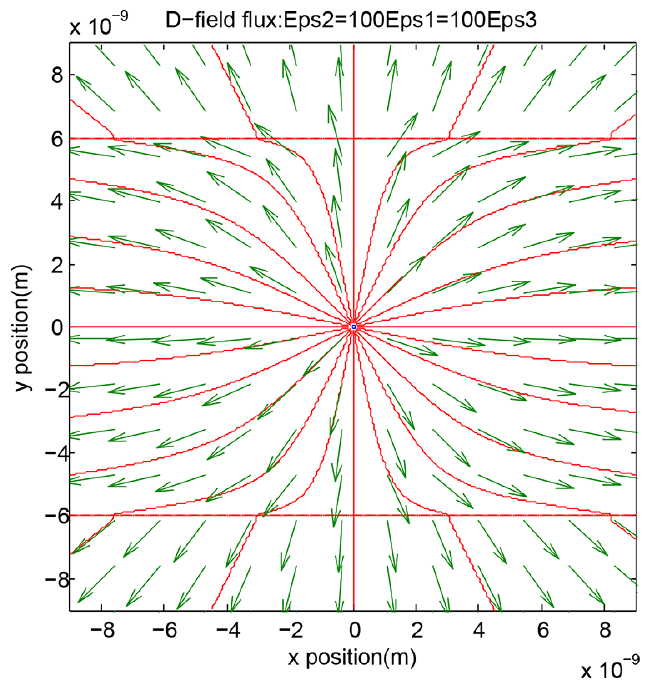}\label{<figure1>}}
     \subfloat[][$\varepsilon_2 \ll \varepsilon_1, \varepsilon_3 $]{\includegraphics[width=0.40\columnwidth]{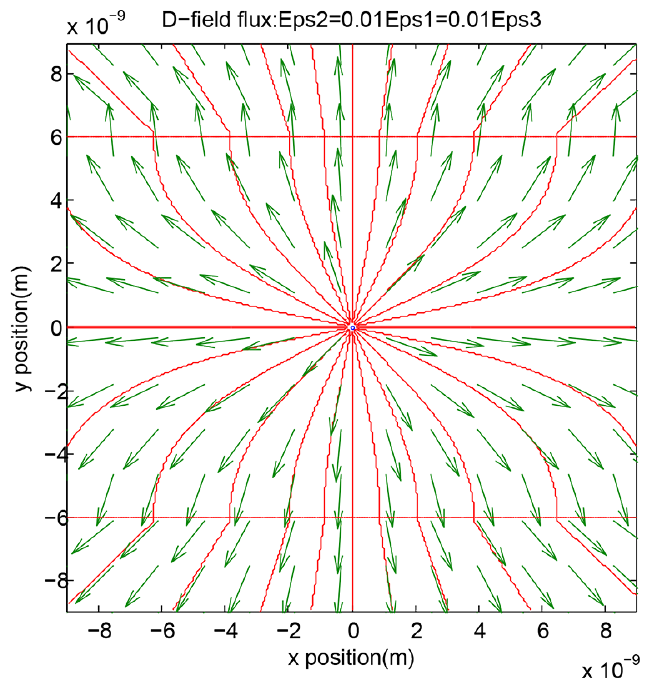}\label{<figure2>}}
     \caption{Lines of electric displacement  in the cross-section of volume $\Omega$ due to a point charge $q$ inside the graphene channel (with electric permittivity $\varepsilon_2$ and is located between top-dielectric ($\varepsilon_3$) and bottom-dielectric ($\varepsilon_1$)) (a) in case $\varepsilon_2 \gg \varepsilon_1, \varepsilon_3 $ and (b) in case $\varepsilon_2 \ll \varepsilon_1, \varepsilon_3 $.}
     \label{electric_field}
\end{figure}

In this preliminary study, we solve the 3D Poisson's equation for a moving electron within an arbitrary three-layer structure of \fref{figure3}, a graphene (with electric permittivity $\varepsilon_2$) channel is located between the first (bottom dielectric, electric permittivity $\varepsilon_1$) and third (top dielectric, electric permittivity $\varepsilon_3$) layers. More details are provided in Appendix A.   The electric displacement lines in the 2D plane X-Y of volume $\Omega$ due to a point charge $q$ inside the graphene channel  are plotted in \fref{electric_field}.

In the case $\varepsilon_2 \gg \varepsilon_1, \varepsilon_3 $, shown in \fref{electric_field}(a), the lines start from the point charge and tend to reach the bottom/top regions. However, the difference on the values of the dielectrics in the three regions implies a tendency of the electric lines to keep in region $ \varepsilon_2$. A moving electron inside the channel has a strong influence on surfaces $S_1$ or $S_4$ (right and left in the figure) and leads to a non-negligible displacement current there. On the contrary, displacement current on surface $S_2$ and surface $S_5$ in volume $\Omega$, top and bottom in the \fref{electric_field}(a), can be somehow neglected. Thus, in conclusion, although dealing with a 3-terminal device, the different dielectrics tend to provide an electric field similar to that found in a 2-terminal device. 

On the contrary, when $\varepsilon_2 \ll \varepsilon_1, \varepsilon_3 $, most lines of the electric field tend to reach surface $S_2$ and surface $S_5$ in volume $\Omega$, top and bottom in the \fref{electric_field}(b), providing a non-negligible current in the gates in a 3-terminal device. Thus, the instantaneous current on $S_4$ do not need to be equal to that on $S_1$, while still satisfying instantaneous current conservation. 

From this simple results, we conclude that a proper engineering design of the different electric permittivity allows to maximize/minimize the displacement current collected on the gates. In next subsection, we will show that this type of manipulation of the source, drain and gate currents can be also realized by modifying the lateral areas of the graphene FETs. 

%%%%%%%%%%%%%%%%%%%%%%%%%%%%%%%%%%%%%%%%%%%%%%%%%%%%%%
%%%%%%%%%%%%%%%%%%%%%%%%%%%%%%%%%%%%%%%%%%%%%%%%%%%%%%
%%%%%%%%%%%%%%%%%%%%%%%%%%%%%%%%%%%%%%%%%%%%%%%%%%%%%%

\subsection{Source, gate and drain time-dependent total current in a 3-terminal device} 
\label{types}

We study now the dependence of the total current (for an electron traversing the graphene transistor) on the device geometry. In order to compute the current due to an electron with a trajectory $\vec{r}(t)=\{ v_x \cdot t, 0, 0 \}$ with velocity $v_x=5 \times 10^5\;m/s$ moving in graphene FETs (the active region is volume $\Omega$), as depicted in \fref{figure3}, we define the time-dependent drain current as $I_D(t)=I_{S_1} + I_{S_3'} + I_{S_6'}$, source current $I_S(t)=I_{S_4} + I_{S_3''} + I_{S_6''}$, gate current $I_G(t)=I_{S_2} + I_{S_5}$. These definitions of the gate, source and drain currents satisfy the requirement of $I_D(t) + I_S(t) + I_G(t) =0$, at any time, because we know that $\sum_{j=1}^{6} I_{S_j}(t) = 0$. Here $S_3'$ and $S_6'$ are the right half parts of $S_3$ and $S_6$, respectively. $S_3''$ and $S_6''$ are respectively the rest part of $S_3$ and $S_6$ to ensure that $S_3' + S_3''=S_3$ (indicated in \fref{figure3}) and $S_6' + S_6''=S_6$. 
\begin{figure}
     \centering
     \subfloat[][Total drain current]{\includegraphics[width=0.50\columnwidth]{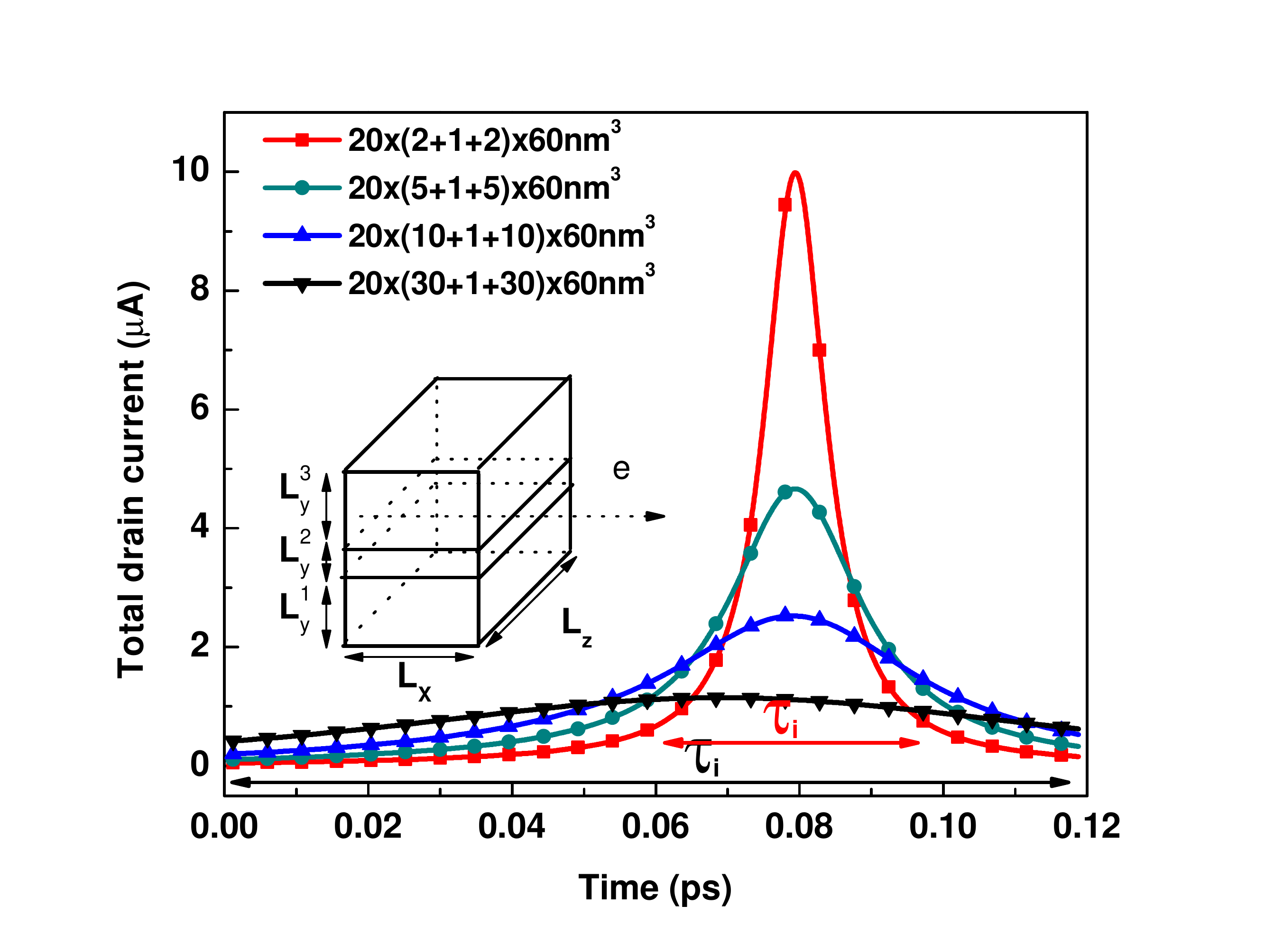}\label{<figure1>}}
     \subfloat[][Total gate current]{\includegraphics[width=0.50\columnwidth]{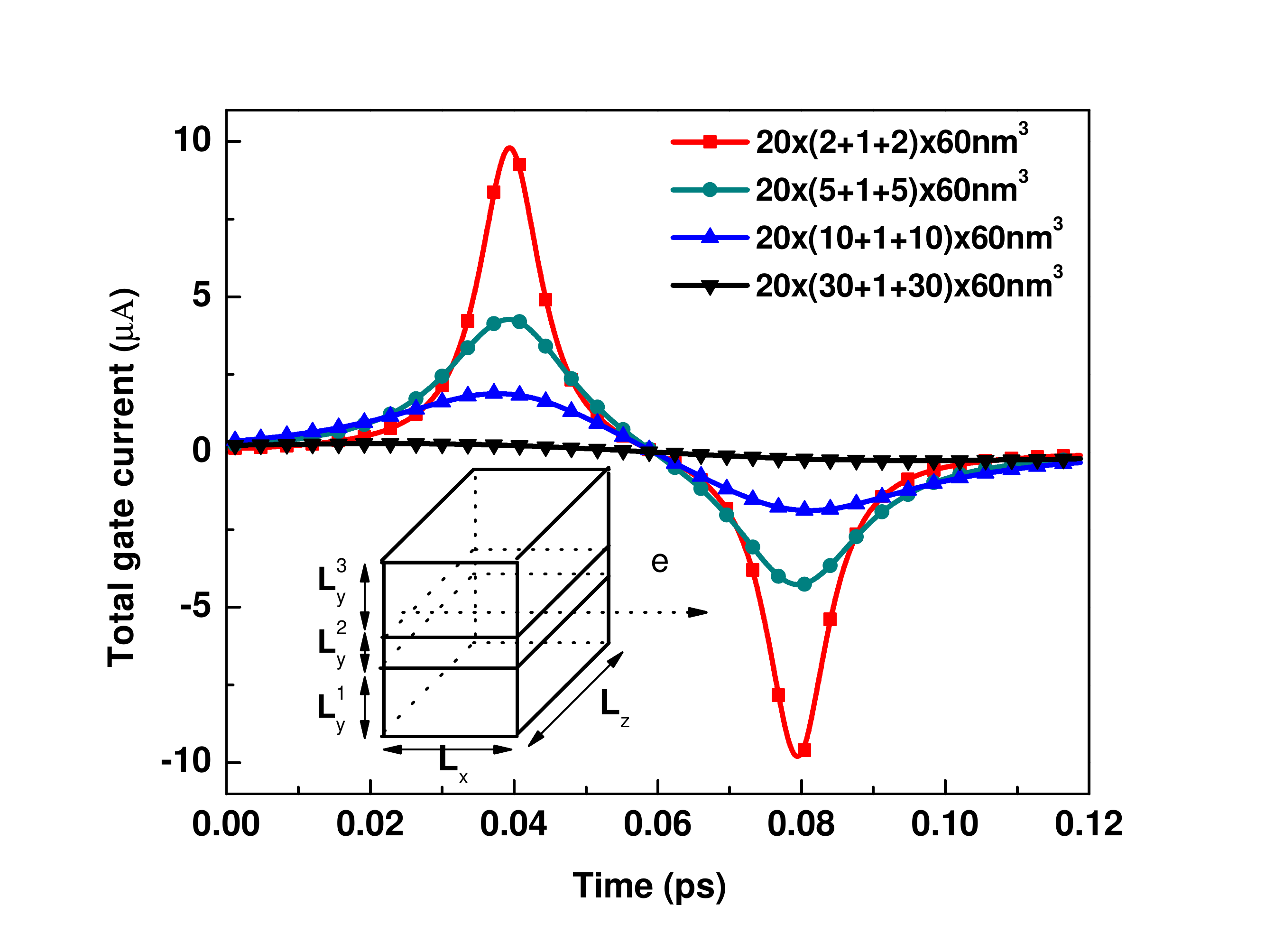}\label{<figure2>}}
     \caption{Total current (a) $I_D(t)$ and (b) $I_G(t)$ due to an electron traversing the volume $\Omega=L_x \times (L_y^1+L_y^2+L_y^3) \times L_z$. A fixed length $L_x=20\;nm$ and several lateral areas by changing height $L_y$ but constant width $L_z=60\;nm$, are considered. }
     \label{height}
\end{figure}

An electron moving inside the channel generates a time-dependent electric field $\vec{E}(\vec{r},t)$ on surface $S_1$, which finally affects the drain current $I_D(t)$.  The value of $I_D(t)$ plotted in \fref{height}(a) shows the dependence of the temporal pulse $\tau_i$ of the current on the lateral area $L_y \times L_z$.  We emphasize that  a unique trajectory, meaning a unique electron transit time $\tau_e=L_x/v_x$, with a fixed length $L_x=20$ nm, is used in all computations where only the lateral dimension $L_z$ is modified. We conclude from \fref{height} that the shape of the current pulse (transient current) is strongly dependent on the lateral surface $L_y \times L_z$. For the structure $20 \times (30+1+30) \times 60\;nm^3$ with $\Omega=L_x \times (L_y^1+L_y^2+L_y^3) \times L_z$, we recover the old result of Ramo \cite{Ramo} and of Shockley \cite {Shockley}, i.e. a current constant in time. Due to the symmetry, the dependence of the source current (not plotted) on the geometry is exactly the same as that for the drain current (with a negative sign).

We observe a quite different behavior for the gate current. We observe two peaks, one positive and one negative, in the gate current. The positive part corresponds to increasing the electric field collected in the gate when the electron is approaching the center of the volume $\Omega$, while the negative values correspond to decreasing the electric field on the gate surface when the electron is leaving. Between positive and negative parts the gate current has to cross the zero. Interestingly, the temporal distance between the maximum and minimum values of the gate current increases for larger lateral areas. 
 
Before providing full numerical results with the BITLLES simulator, let us mention that a lot of information about the frequency spectrum of the current fluctuations can be anticipated from these simple results. On one hand, the Fourier transform of these currents provides direct information on the maximum frequencies of the corresponding spectrum. For example, as a general rule, higher frequencies are required to build sharp peaks associated to the currents of small lateral areas, while lower frequencies are required for the soft peaks associated to  the current with large lateral areas.         
             
%%%%%%%%%%%%%%%%%%%%%%%%%%%%%%%%%%%%%%%%%%%%%%%%%%%%
%%%%%%%%%%%%%%%%%%%%%%%%%%%%%%%%%%%%%%%%%%%%%%%%%%%%
%%%%%%%%%%%%%%%%%%%%%%%%%%%%%%%%%%%%%%%%%%%%%%%%%%%%
\section{Monte Carlo simulation of high-frequency graphene FETs}

We have carried out simulations of graphene FETs using the BITLLES simulator \cite{bitlles} solving the Monte Carlo solutions of the Boltzmann equation (adapted for graphene structures)  to verify the predictions anticipated in the previous section. Before analyzing the numerical results, let us discuss the device under study and the computation of the high-frequency spectrum by post-processing the data of the time-dependent total current.

\subsection{Device definition}

We are simulating the device in \fref{figure3}. An ideal Ohmic contact is assumed in the drain and source contacts to ensure that the applied $V_{DS}$ directly translates into a difference of the source and drain Fermi levels, i.e. $E_{fd}=E_{fs}-qV_{DS}$, being $E_{fs}$ and $E_{fd}$ the source and drain Fermi levels, respectively. The extrinsic role of the contact resistance mentioned in the introduction is directly disregarded in all the intrinsic results discussed in this work. In particular, we consider $E_{fs}=0.05\;eV$ with $V_{TG}=V_{BG}=0.05\;V$ (being $V_{TG}$ top gate voltage and $V_{BG}$ bottom gate voltage), $V_S=0\;V$ and $V_D=0.1\;V$.

\subsection{Klein tunneling, band-structure and positive-negative energy injection}

Due to the 2D nature of graphene, only two degrees of freedom specify the electron position, $\{x,z\}$. Equivalently, only two wave vectors $\{k_x,k_z\}$ are needed. As a consequence of the honeycomb graphene structure, the relationship between the energy of electrons, $E_k$ and its wave vector $|\mathbf{k}|=\sqrt{k_x^2+k_z^2}$ is: 
\begin{equation}
E_k=\pm \hbar v_{f}|\mathbf{k}|,
\label{ek}
\end{equation}
 where $v_f=5\times10^5\;m/s$ is the Fermi velocity. This linear $E_k-k$ dispersion has several relevant differences with typical  parabolic $E_k-k$ dispersion in Silicon that, at the end of the day, implies important differences with the typical Monte Carlo tools:
\begin{itemize}
\item Electrons with positive and negative (kinetic) energy in a gapless material: From the $\pm$ signs in equation \eref{ek}, we notice that there are two possible energies for an electron with momentum $|\mathbf{k}|$. In the literature, usually those electrons with negative energy are called holes and with positive energy called electrons. We do not use the name hole because here we simulate explicitly electrons in the conduction band (CB, positive kinetic energy) and electrons in the valence band (VB, negative kinetic energy). The graphene (kinetic) energy band structure has a zero (kinetic) energy at the Dirac point. Contrary to Silicon semiconductors, there is no energy gap between electrons with positive energy and electrons with negative energy.  Injections from the contact have to consider electrons above and below the Dirac point, as illustrated in \fref{Injection}(a). The details of the particular positive-negative (kinetic) energy injection model in the BITLLES simulator will be explained somewhere else. 

\begin{figure}
     \centering
     \subfloat[][Positive-negative energy injection]{\includegraphics[width=0.50\columnwidth]{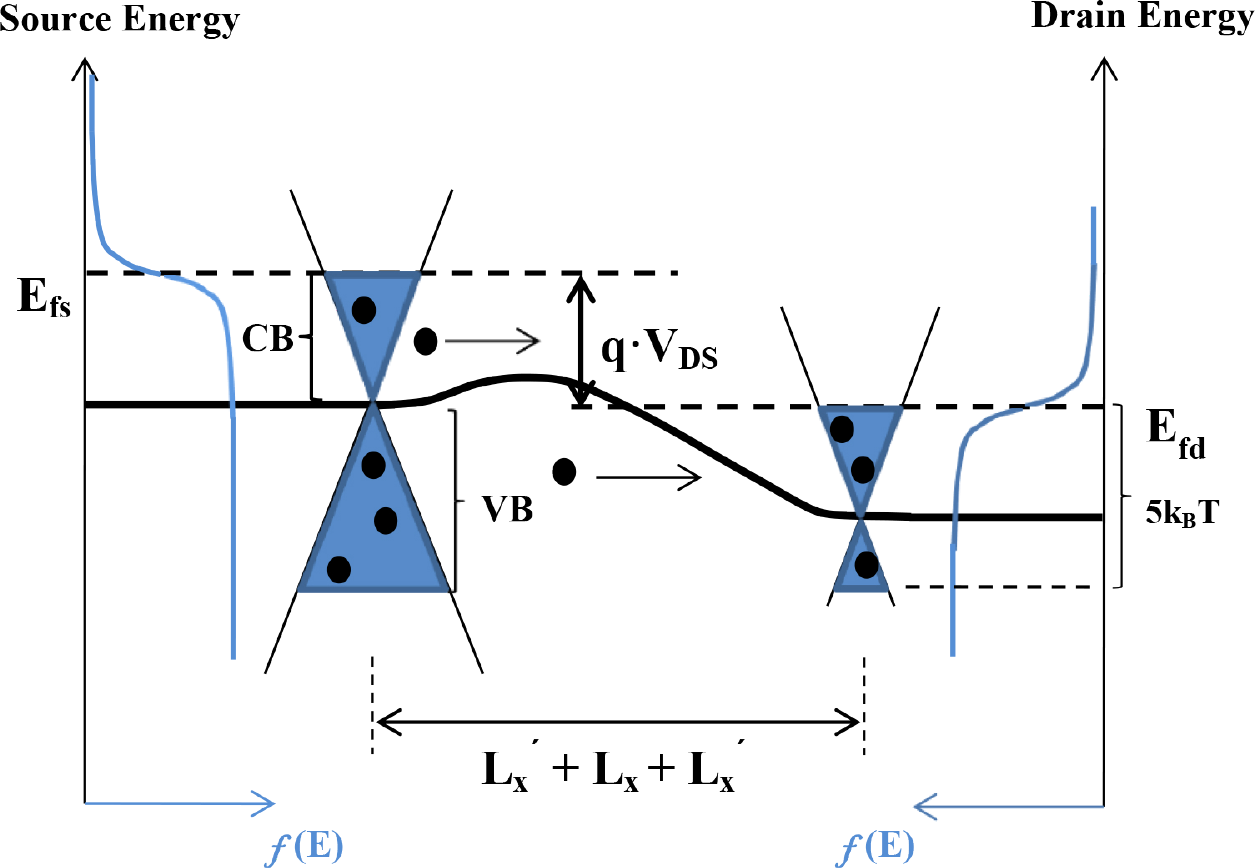}\label{electron-hole}}
     \subfloat[][Trajectory]{\includegraphics[width=0.50\columnwidth]{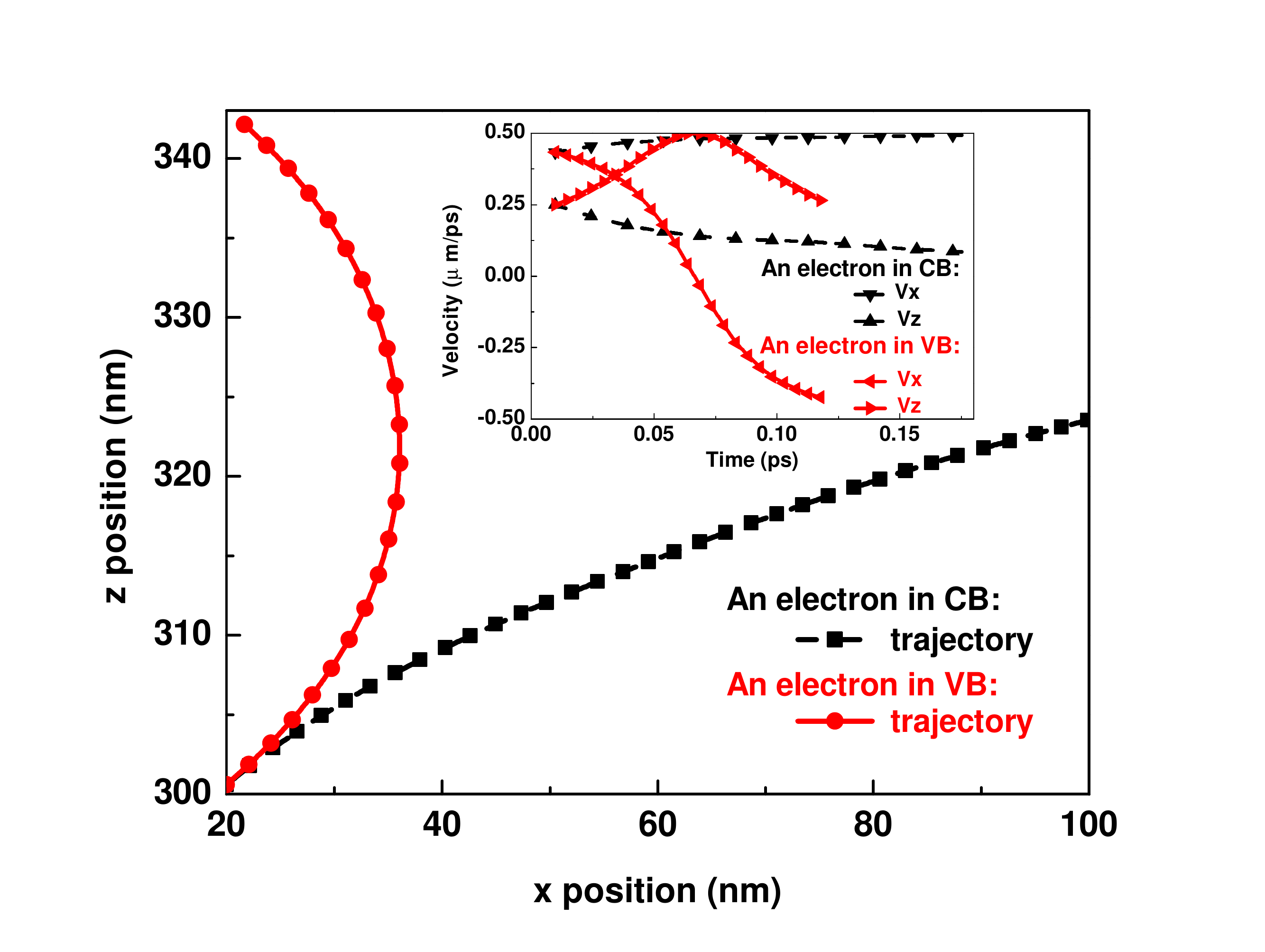}\label{trajectory}}
     \caption{(a) Schematic representation of an energy profile in the transport direction of graphene transistors in BITLLES simulator. The applied drain-source bias, $V_{DS}$, provides a different source, $E_{fs}$, and drain, $E_{fd}$,  Fermi levels, the Fermi-Dirac distribution function at each contact are indicated. (b) The trajectory for one electron above the Dirac point and one below with the same absolute kinetic energy injected from the source approaching the drain.}
     \label{Injection}
\end{figure}

\item Transport equation in graphene: From equation \eref{ek} one can easily find the semi-classical equations of motion for electrons in graphene. Their dynamical behavior is similar to massless relativistic particles. However, their maximum velocity is not the speed of light, but $v_f$. A simple understanding of graphene electrons under an applied bias (from drain to source) can be achieved by using the conservation of the total energy $E$ and of the momentum $k_z$ in the $z$ direction (no applied bias in that direction). If we define $\theta$ as the angle between $\mathbf{k}$ and $k_x$, i.e. $sin(\theta)=k_z/|\mathbf{k}|$, then the conservation of $E$ and of z-momentum imply that an electron of total energy $E$ moving from a location with potential energy $U_o$ till another point with $U_f$ has to satisfy the relation:
\begin{equation}
(E-U_o) sin(\theta_o) = \pm (E-U_f) sin(\theta_f) 
\label{para}
\end{equation}
When an electron with positive kinetic energy $E_k$ moves with $U_o>U_f$ (and with both $U_o<E$ and $U_f<E$), then, it tends to reach the maximum velocity in the $x$ direction and minimum in the $z$ direction (dashed lines in the inset of \fref{Injection}(b)) and the contrary if $U_o<U_f$. The opposite behavior (solid lines in the inset of \fref{Injection}(b)) is found for an electron with negative kinetic energy $E_k$ (with both $U_o>E$ and $U_f>E$). All previous trajectories are compatible with using the sign $+$  in equation \eref{para}. On the contrary, if the electron trajectory involves changing from positive to negative kinetic energies because of Klein tunneling (either $U_o<E$ and $U_f>E$ or $U_o>E$ and $U_f<E$), then, the sign $-$  has to be used in equation \eref{para}.

\item Because of the energies available above and below the Fermi point, the typical gap barrier with forbidden energies found in Silicon is unachievable in graphene. This exotic phenomenon leads to a particular effect called Klein tunneling \cite{Novoselov2006, Jena}. When an electron impinges with sharp variations of the potential energies,  there is a large probability that the electron tunnels from one band to another \cite{Novoselov2006, Jena}. This kind of band-to-band tunneling implies that the tunneling current is always relevant, resulting in a failure to get saturation near the pinchoff \cite{David, Paussa}. The tunneling probability in the $x$ direction is given by \cite{Zhang}:
\begin{equation}
 T = exp\Big( -\pi \hbar v_f k_z^2/(e|F|)  \Big), 
 \label{kleintun}
\end{equation}
 where $|F|$ is the magnitude of the local electric field in the $x$ direction, and $k_z$ is the wave vector in $z$ component. Similar probability expression is used for the tunneling in the $z$ direction. 
\end{itemize}

\subsection{Simulation definition}

All simulations of the device in \fref{figure3} are performed with a DC polarization for the gate, source and drain bias. In any case, since we deal with an explicit time-dependent formalism, we will capture the intrinsic dynamics of the electrons in the active device region. Each bias point is simulated during $T=1000\;ps$, with a time step of $\Delta t=0.7\;fs$. The Poisson equation is solved in the whole active device region depicted in \fref{figure3} including $L_x'$, larger than $\Omega$, with Dirichlet boundary conditions on the gates, source and drain surfaces and Dirichlet in the rest of boundaries. At each time step of the Monte Carlo simulation, $\Delta t$, the total current $I(t)$ in all surfaces of the volume $\Omega$ of \fref{figure3} are computed, following equations \eref{displacement} and \eref{particle}. From these currents on the six faces of the volume $\Omega$, the source, gate and drain currents are computed following the definitions given in \sref{types}. Then, the frequency spectrum of current fluctuations is computed. The power spectral density (PSD) can be defined as the Fourier transform of the time-average definition of the autocorrelation function $\Delta R(\tau)$ \cite{Oriols2001, Marian}:
\begin{equation}
\label{auto}
\Delta R(\tau)=\lim_{T \to \infty} \frac{1}{T} \int_0^T I(t)I(t+\tau) dt -\langle I \rangle ^2,
\end{equation}
where the current $I(t)$ can be measured at the gate, source or drain. The value $\langle I \rangle$ is the time average value of current $I(t)$ during a large period of time $T$:
\begin{equation}
\label{average_current}
\langle I \rangle=\lim_{T \to \infty} \frac{1}{T} \int_0^T I(t)dt.
\end{equation}
Then, the Fourier transform of equation \eref{auto} gives the PSD of the noise $S(\omega)$:
\begin{equation}
\label{PSD}
S(\omega)= \int_{-\infty}^{\infty} \Delta R(\tau)e^{-j\omega \tau} d\tau.
\end{equation}

%%%%%%%%%%%%%%%%%%%%%%%%%%%%%%%%%%%%%%%%%%%%%%%%%%%
%%%%%%%%%%%%%%%%%%%%%%%%%%%%%%%%%%%%%%%%%%%%%%%%%%%
%%%%%%%%%%%%%%%%%%%%%%%%%%%%%%%%%%%%%%%%%%%%%%%%%%%
\section{Numerical results}

We consider two different geometries for the graphene FET depicted in \fref{figure3}. We define device A with  a volume $\Omega_A=20 \times (2+1+2) \times 60\;nm^3$ and device B is $\Omega_B=20 \times (30+1+30) \times 60 \;nm^3$ with $\Omega=L_x \times (L_y^1+L_y^2+L_y^3) \times L_z$. In both geometries, the length in the transport direction is $L_x=20\;nm$ (identical transit times), the height in the $y$ direction of the 2D graphene sheet is $1\;nm$  and width in the $z$ direction is  $L_z=60\;nm$. The only difference is the height of the dielectrics drawn in \fref{figure3}. All the rest physical parameters discussed in the previous section are identical for both structures.

In order to test the effect of the Klein tunneling and positive-negative (kinetic) energy injection discussed in the previous section, we consider high-frequency performance for three types of simulations for both structures. The results of the DC currents and low-frequency noise with or without Klein tunneling and with or without negative energy injection are summarized in table \ref{tabla}.

\subsection{PSD without Klein tunneling and only positive energy injection }

\begin{figure}[h!!!]
\centering
\includegraphics[width=0.80\columnwidth]{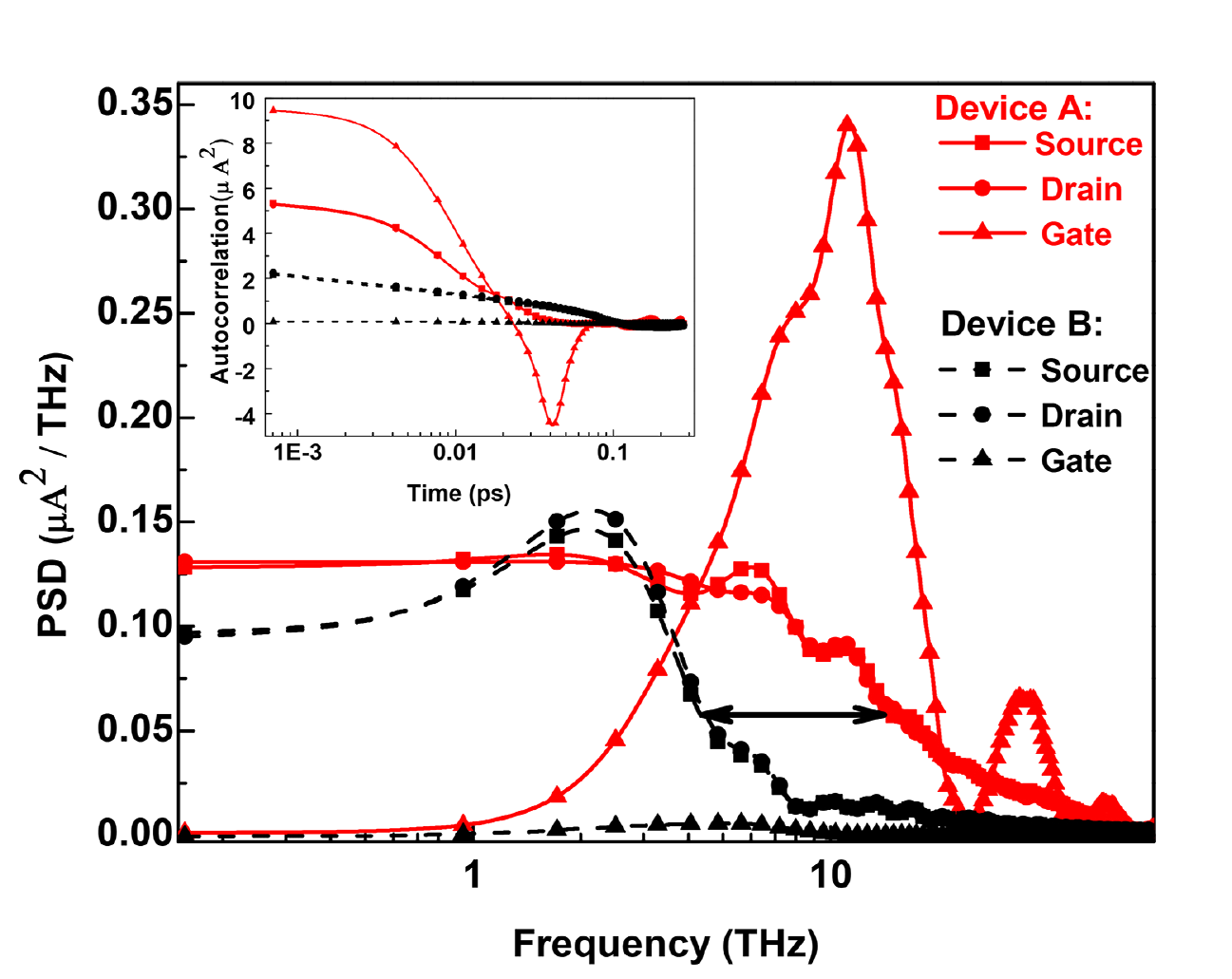}
\caption{PSD of the current fluctuations as a function of frequency for graphene FETs with two different geometries (but identical channel length $L_x = 20\;nm$) in device A and B, operating under DC conditions: double gates $V_{TG}=V_{BG}=0.05\;V$, applied bias $V_{DS}=0.1\;V$, without Klein tunneling and only positive kinetic energy injection.}
\label{figure6}
\end{figure}
First of all, we consider a simulation without Klein-tunneling and only positive kinetic energy injection. This is a type of simulation similar to the one done for Silicon FETs.  In \fref{figure6}, we plot the PSD of current fluctuations for the two different geometries. First, we observe that the (maximum) frequencies where the PSD of the drain and source current fluctuations drops down to zero become different and with a difference of almost one order of the magnitude. Let us emphasize that, in principle, both geometries in \fref{figure6} have roughly the same electron transit time $\tau_e=L_x/v_x$. A displacement of the noise spectrum to the higher frequency range can be achieved without changing the device active region $L_x=20\;nm$ nor its (average) velocity $v_x$.  The physical reason of this effect can be easily understood from the results of \fref{height} for the drain and source currents and the explanation there. Sharp temporal peaks of the displacement currents requires higher frequencies, and vice versa. This very relevant effect will also appear in all the rest of simulations and its consequences in the cut-off frequencies will be mentioned in the conclusions. 

In addition, in \fref{figure6} we observe a large peak of the gate current for device A at frequency $f=10\;THz$, much larger than that of device B. This effect can also be clearly seen in the autocorrelation plotted in the inset of the figure. The sample B has a much lower value of the PSD of the gate current peak and its peak appears at a lower frequency $f=4\;Thz$. The basic features of these peaks of the gate current again can be straightforwardly understood from the results of \fref{height}. The Fourier transform of the gate current drawn in \fref{height} is basically a delta in the frequency of the oscillatory signal for device B and several deltas in the case of sample A. Notice the tendency to a multi peak spectrum in the PSD of device A. In both cases, no PSD appears at zero frequency because there is no particle current in the gate, but only displacement current that goes to zero when averaged in a long period of time. 

\begin{figure}[h!!!]
\centering
\includegraphics[width=0.80\columnwidth]{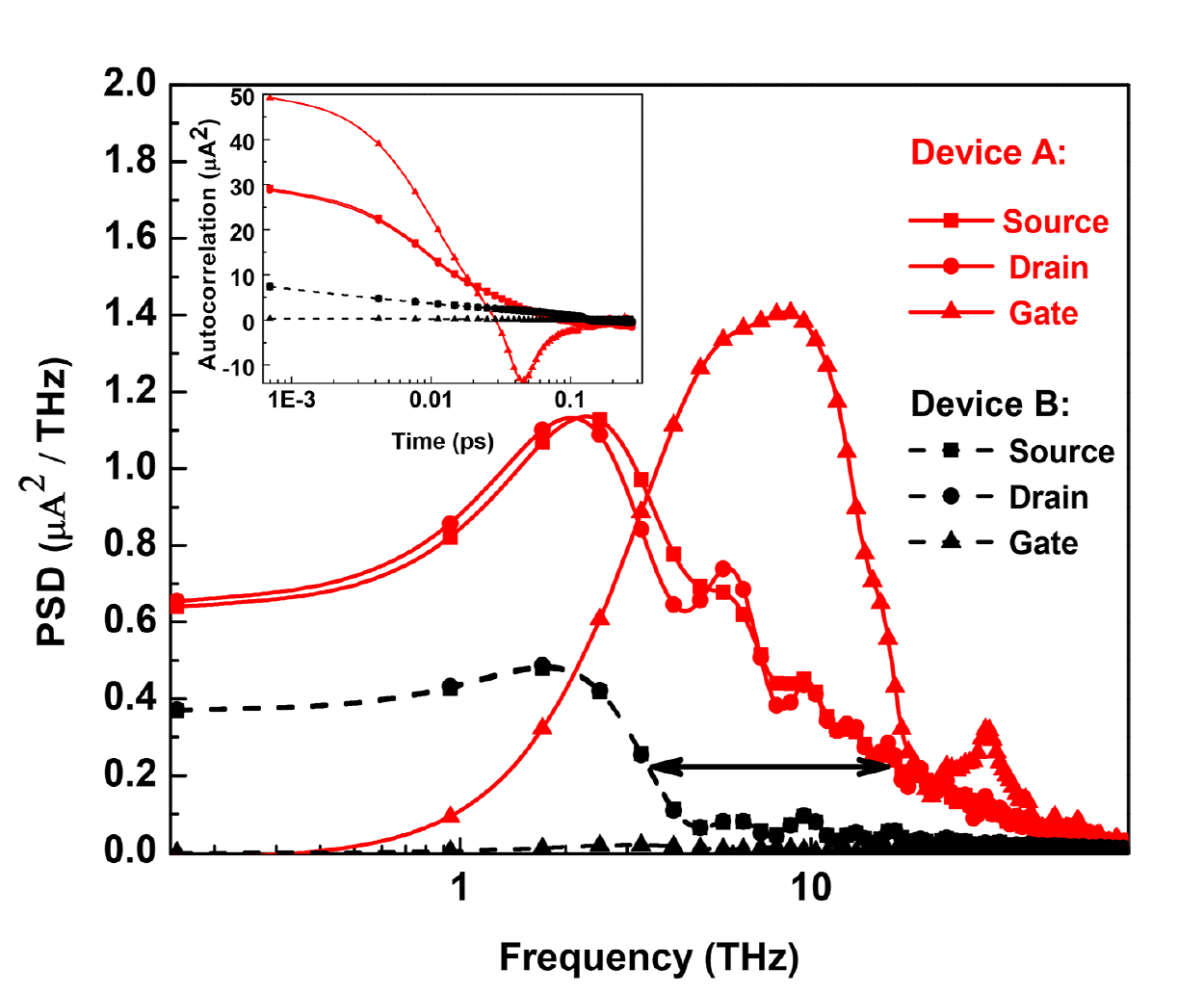}
\caption{PSD of the current fluctuations as a function of frequency for graphene FETs with two different geometries (but identical channel length $L_x = 20\;nm$) in device A and B, operating under DC conditions: double gates $V_{TG}=V_{BG}=0.05\;V$, applied bias $V_{DS}=0.1\;V$, with Klein tunneling and only electron injection from above the Dirac point.}
\label{figure7}
\end{figure}

\subsection{PSD with Klein tunneling and only positive energy injection }

When the Klein tunneling is considered, but still only electron injection from above the Fermi Dirac point, the results in \fref{figure7} are qualitatively very similar to the ones plotted in \fref{figure6}. However, let us emphasize that the PSD is basically one order of magnitude larger now as written in table \ref{tabla}. The Klein tunneling with a \lq\lq{}random\rq\rq{} tunneling probability of being reflected or transmitted given by equation \eref{kleintun} (with the change from positive kinetic energy to negative kinetic energy or vice versa) introduces an important source of noise. These new source of noise cannot be avoided in graphene and is present even in the \rq\rq{}ballistic\rq\rq{} regime (no phonon or impurity scattering) considered  in this work.   

Looking at the DC currents in table \ref{tabla}, we realize that this noise increment is obtained, in fact, with a reduction of the DC current (the signal). The physical reason of the reduction is quite simple. Now, electrons from the drain are able to reach the source, even with an applied bias $V_{DS}=0.1\;V$, because of the Klein tunneling. This new addition of flux of electrons has an opposite sign when compared to the conventional source-to-drain current. Certainly, the signal-to-noise ratio is greatly degraded because of the Klein tunneling. This is an important and unavoidable drawback for high-frequency applications of graphene, not usually noticed in the literature.

\begin{figure}[h!!!]
\centering
\includegraphics[width=0.80\columnwidth]{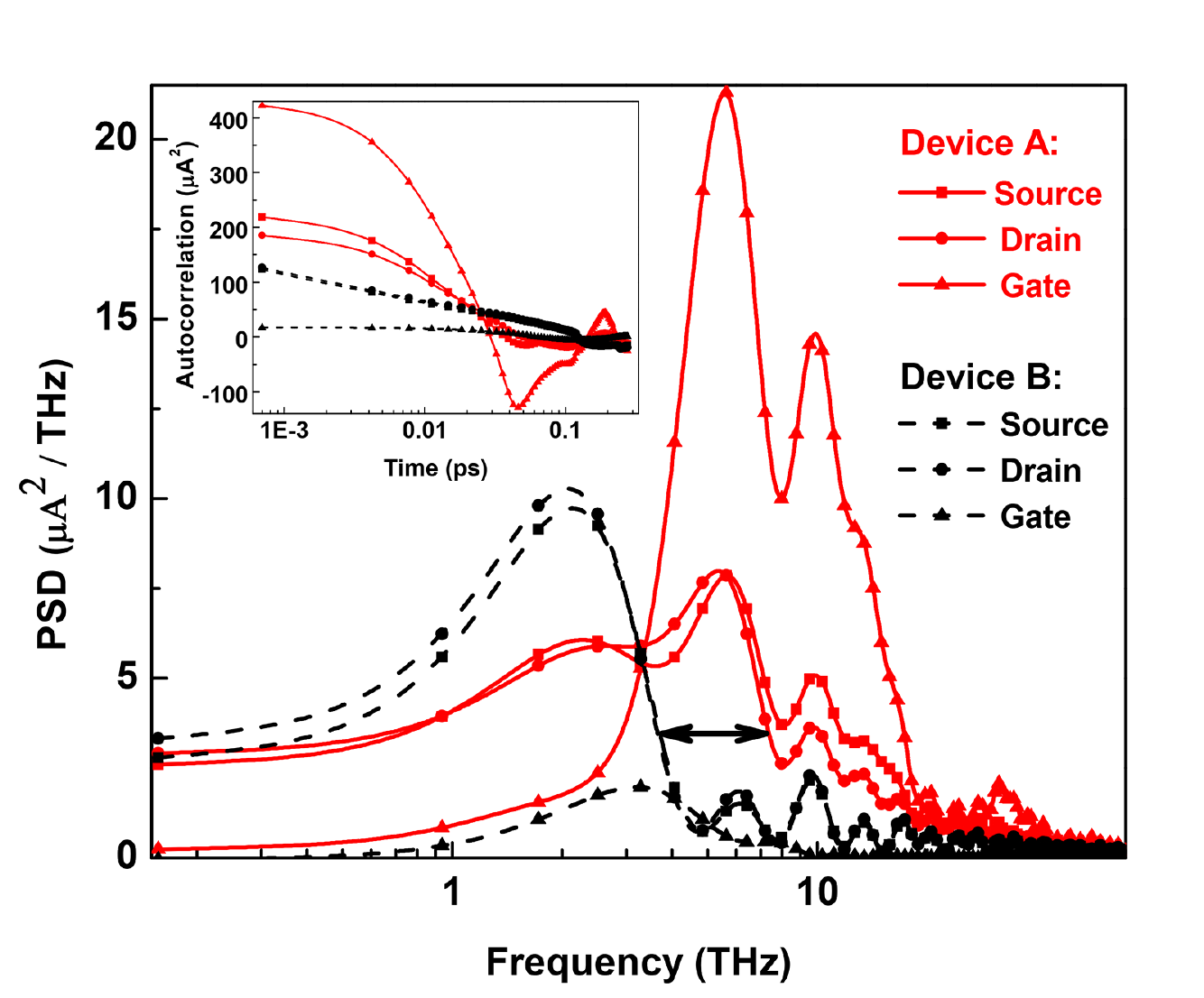}
\caption{PSD of the current fluctuations as a function of frequency for graphene FETs with two different geometries (but identical channel length $L_x = 20\;nm$) in device A and B, operating under DC conditions: double gates $V_{TG}=V_{BG}=0.05\;V$ , applied bias $V_{DS}=0.1\;V$, with Klein tunneling and positive-negative energy injection.}
\label{figure8}
\end{figure}
      
\subsection{PSD with Klein tunneling and positive-negative energy injection   }

Finally, in \fref{figure8}, we plot the PSD for device A and B when, both, Klein tunneling and positive-negative energy injection are considered. The main features discussed for \fref{figure6} and \fref{figure7} are also present in these new results. Again we emphasize that the PSD increases two orders of magnitude, while the DC current is roughly a factor of 10 greater than that in the first case. For the noise, the new thermal injections from below the Fermi point implies larger noise. For the DC current, now, there is a combination of a decrement of the DC current due to Klein tunneling and an increment due to the source-to-drain injection from below of the Fermi Dirac point (greater than the drain-to-source injection). Let us mentioned that we do consider the effect of the Pauli principle in the injection model, but we do not consider it during the dynamics of electrons in the device active region. Therefore, the DC current is a little overestimated. In any case, the important degradation of the signal-to-noise ratio mentioned for the results of \fref{figure7} is also present here.   

\begin{table}[htb]
\begin{center}
\caption{DC current and zero-frequency noise (i.e. $S(\omega \to 0)$) for three types of simulations for both device A and device B. In the table, KT means simulation with Klein tunneling,  PI with positive (kinetic) energy injection and NI with negative kinetic energy injection.}
\begin{tabular}{|c|c|c|c|c|c|c|}
\hline
  \multicolumn{3}{|c|}{ }  &\multicolumn{2}{c|}{Device A }                      &\multicolumn{2}{c|}{Device B }\\ \hline
    \multicolumn{3}{|c|}{ }&DC current($\mu$A)&noise($\mu A^2$/THz) &DC current($\mu$A) &noise($\mu A^2$/THz)\\ \hline
     & PI  &                & 0.68                        &0.130                            &  1.50                        & 0.095\\ \hline
KT & PI  &                         & 0.21                        &0.641                            &  0.31                        & 0.368 \\ \hline
KT & PI  &NI                     & 10.2                        &2.711                            &  8.14                        & 2.956 \\ \hline
\end{tabular}\\
\label{tabla}
\end{center}
\end{table}

%%%%%%%%%%%%%%%%%%%%%%%%%%%%%%%%%%%%%%%%%%%%%%%%%%%%%%%
%%%%%%%%%%%%%%%%%%%%%%%%%%%%%%%%%%%%%%%%%%%%%%%%%%%%%%%
%%%%%%%%%%%%%%%%%%%%%%%%%%%%%%%%%%%%%%%%%%%%%%%%%%%%%%%
\section{Conclusions}

In this work, we open a new path to study the announced THz behavior of graphene transistors. Instead of using the extended strategy of providing high-frequency predictions from quasi-static simulations, we directly simulate the time-dependent particle and displacement currents in an intrinsic graphene FET .  We just simulate explicitly the device active region, thus, all extrinsic effects due to (parasitic) resistances in the source, drain or gate contacts are directly ignored in this work. We only focus on the intrinsic high-frequency effects due to the exotic gapless linear (energy-wave vector) band structure of graphene. We use a Monte Carlo solution of the Boltzmann equation fully adapted to graphene FETs. In particular, the semi-classical transport equations for electrons above and below the Dirac (zero energy) point are adapted accordingly. In addition, the Klein tunneling is explicitly considered allowing electrons to transit from above to below the Dirac point, or vice versa. A novel electron injection model for electrons with positive and negative kinetic energies are developed for graphene (including the thermal noise and Fermi Dirac statistic). From the simulations of the dynamics of the electrons inside the active region (under constant DC polarization in the gate, source and drain contact), we compute the \emph{measured} total currents in the three FET terminals. From such time-dependent currents, the PSD of their fluctuations is computed as a Fourier transform of the current autocorrelation.  

Particular features of high frequency behavior of graphene FETs are predicted from such PSD.  We perform  simulation with and without Klein tunneling, and injection from positive or both positive and negative (graphene injection) kinetic energies. From such simulations, we conclude that the unavoidable Klein tunneling and graphene injection provide an increment of noise at THz frequencies (and also at lower frequencies) when compared to simulations without Klein tunneling or graphene injection. Such increment of the noise is not compensated by a similar increment on the average DC current (interpreted here as the signal), providing an unavoidable degradation of the signal-to-noise ratio. Certainly, the use of a semi-classical simulation tool is an approximation, however, we have already tested with (Bohmian) quantum solutions of the Dirac equation that the extension towards quantum simulations tools will not provide important variations \cite{Oriols2007, Oriols2009, Albareda, Oriols2013}. The main approximation in this present work (common in most Monte Carlo simulations) is the fact that the Pauli principle (the exchange interaction between electrons) is not explicitly considered during the dynamics of electrons in the device active region. 

The two geometries of the graphene FET studied in this work, device A and device B, have exactly the same length in the $x$ (drain-source transport) direction. Therefore, both geometries imply the same transit time, $\tau_e$, but they have different temporal width of the peak current $\tau_i$, see \fref{height}. However, the shorter the vertical height (in comparison with the length of the active region in the transport direction), the larger the maximum frequency of the PSD. This can be seen in \fref{figure6}, \fref{figure7} and  \fref{figure8}.  From this result we can envision an alternative strategy (without length scaling) to optimize the intrinsic cut-off frequency of graphene transistors. It is argued from the usual (quasi-static approximation) predictions of the cut-off frequency, $f_T$, that its value is inversely proportional to the transit time, $f_T \approx 1/(2 \pi \tau_e)$,  pointing out that the electron transit time as the ultimate limiting factor \cite{Schwierz2010, Howes}. This last result suggests to improve the material mobility and shorten the $x$ transport direction when optimizing $f_T$. However, in our work, we have shown that in fact a careful time-dependent analysis of the displacement current generated by a moving electron shows that, for some particular graphene FETs named here as device A (with lateral dimensions much shorter than their gate length), the limiting effective time is the current peak, $\tau_i$, which can be much smaller than $\tau_e$, This can be seen in \fref{figure6},  \fref{figure7} and  \fref{figure8}. The same result can be anticipated from \fref{height}.  

Future work will be devoted to the analysis of the cut-off frequency computed from the explicit simulation of the time-dependent displacement and particle currents under small-signal (AC polarization for the drain and gate contacts) conditions in graphene FETs with either semi-classical trajectories or (Bohmian) quantum trajectories solution of the Dirac equation. 

%%%%%%%%%%%%%%%%%%%%%%%%%%%%%%%%%%%%%%%%%%%%%%%%%%%%%%%
%%%%%%%%%%%%%%%%%%%%%%%%%%%%%%%%%%%%%%%%%%%%%%%%%%%%%%%
%%%%%%%%%%%%%%%%%%%%%%%%%%%%%%%%%%%%%%%%%%%%%%%%%%%%%%%
\section*{Acknowledgement}
\addcontentsline{toc}{section}{Acknowledgement}
We acknowledge support from the \lq\lq{}Ministerio de Ciencia e Innovaci\'{o}n\rq\rq{} through the Spanish Project TEC2012-31330, Generalitat de Catalunya (2014 SGR-384) and the Grant agreement no: 604391 of the Flagship initiative  \lq\lq{}Graphene-Based Revolutions in ICT and Beyond\rq\rq{}. Z. Zhan acknowledges financial support from the China Scholarship Council (CSC).

%%%%%%%%%%%%%%%%%%%%%%%%%%%%%%%%%%%%%%%%%%%%%%%%%%%%%%%
%%%%%%%%%%%%%%%%%%%%%%%%%%%%%%%%%%%%%%%%%%%%%%%%%%%%%%%
%%%%%%%%%%%%%%%%%%%%%%%%%%%%%%%%%%%%%%%%%%%%%%%%%%%%%%%
\appendix
\addcontentsline{toc}{section}{Appendix A}
\addtocontents{toc}{\protect\setcounter{tocdepth}{-1}} 
\label{three_dielectrics}
\section{Using the image method to solve the problem of a point charge in presence of a three-dielectric-medium with planar interfaces}

The image method described in any physical textbook is used for computing the potential and the electrical distribution around electrostatic charges in the presence of conductors or dielectrics. In the case of a point charge $q$ near a conducting plane, by using the boundary conditions, we can directly write the field due to $q$ and to an imaginary point charge $-q$ at a suitable position. In this section, we will extend the method of the images to the case of an electrostatic point charge in the presence of three arbitrary different dielectric media, i.e. two semi infinite media separated by a sheet.

The \fref{figure3} shows a geometry of the problem: a point charge is imbeded in a three-dielectric medium with infinite planar interfaces. Without loss of generality, we locate the point charge $q$ at an arbitrary point $A_1(x_0,y_0,z_0)$ in the second layer. For simplicity, we consider the situation of planar interfaces perpendicular to the y axis and characterized by the sequence of dielectric constants:

\numparts
\begin{eqnarray}
\label{dielectric_constant}
&\varepsilon_1, \qquad  \qquad \; y\leq Y_1 \\
& \varepsilon_2, \qquad  Y_1 \leq y \leq Y_2\\
&\varepsilon_3, \qquad \qquad \;  y\geq Y_2 
\end{eqnarray}
\endnumparts
Using the method of images \cite{image1, image2}, a straightforward calculation of potential $\Phi$ at an arbitrary point $P(x, y, z)$ described by rectangular coordanates is:

\numparts
\begin{eqnarray}
\label{potential}
\fl \qquad \Phi_1&= \frac{T_{12}q}{4\pi \varepsilon_1}\Big[ \frac{1}{r_0}+\displaystyle\sum_{n=1}^{\infty}(L_{12}L_{32})^{n-1} \Big( \frac{L_{12}L_{32}}{r_{na}^{+}} + \frac{L_{32}}{r_{nb}^{+}} \Big)  \Big]  , \qquad \qquad \qquad \; \; y\leq Y_1 \\
\fl \qquad \Phi_2&=  \frac{q}{4\pi \varepsilon_2}\Big[ \frac{1}{r_0}+\displaystyle\sum_{n=1}^{\infty}(L_{12}L_{32})^{n} \Big(- \frac{L_{32}}{r_{na}^{-}} - \frac{1}{r_{nb}^{-}}-\frac{1}{r_{nb}^{+}}-\frac{L_{12}}{r_{na}^{+}}  \Big)  \Big]  ,  \; Y_1 \leq y \leq Y_2\\
\fl \qquad \Phi_3&= \frac{T_{32}q}{4\pi \varepsilon_3}\Big[ \frac{1}{r_0}+\displaystyle\sum_{n=1}^{\infty}(L_{12}L_{32})^{n-1} \Big( \frac{L_{12}L_{32}}{r_{na}^{-}} + \frac{L_{12}}{r_{nb}^{-}} \Big)  \Big]  , \qquad \qquad \qquad \; \;  y\geq Y_2 
\end{eqnarray}
\endnumparts
where $T_{i2}=\frac{2\varepsilon_i}{\varepsilon_i+\varepsilon_2}$ and $L_{i2}=\frac{\varepsilon_2-\varepsilon_i}{\varepsilon_2+\varepsilon_i}$ with $i=1,3$. The distances between point $P$ and the infinity array of charges are given by:

\numparts
\begin{eqnarray}
\label{distance}
\fl \qquad  r_0&=[(x-x_0)^2 +(y-y_0)^2 +(z-z_0)^2]^{\frac{1}{2}}\\
\fl \qquad  r_{na}^{-}&= [(x-x_0)^2 +(y-y_0+2n(Y_2-Y_1))^2 +(z-z_0)^2]^{\frac{1}{2}} \\
\fl \qquad  r_{nb}^{-}&=   [(x-x_0)^2 +(y+y_0-2Y_1+2(n-1)(Y_2-Y_1))^2 +(z-z_0)^2]^{\frac{1}{2}} \\
\fl \qquad  r_{nb}^{+}&=  [(x-x_0)^2 +(y+y_0-2Y_2-2(n-1)(Y_2-Y_1))^2 +(z-z_0)^2]^{\frac{1}{2}} \\
\fl \qquad  r_{na}^{+}&= [(x-x_0)^2 +(y-y_0-2n(Y_2-Y_1))^2 +(z-z_0)^2]^{\frac{1}{2}} 
\end{eqnarray}
\endnumparts
It has been proved that the infinite series in the potential expressions can be trunkated and that the number of terms $10\sim 20$ is sufficient enough to achieve a good precision \cite{image2}. In BITLLES, within reasonable limits for a rapid calculation, we define $n=100$.  

%%%%%%%%%%%%%%%%%%%%%%%%%%%%%%%%%%%%%%%%%%%%%%%%%%%%%%%%
%%%%%%%%%%%%%%%%%%%%%%%%%%%%%%%%%%%%%%%%%%%%%%%%%%%%%%%%
%%%%%%%%%%%%%%%%%%%%%%%%%%%%%%%%%%%%%%%%%%%%%%%%%%%%%%%%


\section*{References}

\begin{thebibliography}{100}

\bibitem{Ferry2008}
Ferry D K, Gilbert M J and Akis R 2008  {\it IEEE Trans. Electron Devices} {\bf 55} 2820-6

\bibitem{Novoselov2009}
Castro Neto A H, Guinea F, Peres N M R, Novoselov K S and Geim A K 2009 {\it Rev. Mod. Phys.} {\bf 81} 109

\bibitem{Schwierz2010}
Schwierz F 2010 {\it Nat. Nanotechnol.} {\bf 5} 487-496

\bibitem{Geim}
Geim A K 2009 {\it Science} {\bf 324} 1530-4

\bibitem{Schwierz2011}
Schwierz F 2011 {\it Nature} {\bf 472} 41-42

\bibitem{Sordan}
Sordan R and Ferrari A C 2013 {\it IEEE Int. Electron Devices Meeting (Washington)}  pp~1.1.1 - 7

\bibitem{Meric}
Meric I, Baklitskaya N, Kim P and Shepard K L 2008 {\it IEEE Int. Electron Devices Meeting (San Francisco)}  pp~1-4

\bibitem{Lin}
Lin Y M, Dimitrakopoulos C, Jenkins K A, Farmer D B, Chiu H Y, Grill A and Avouris P 2010 {\it Science} {\bf 327} 662

\bibitem{Liao}
Liao L, Lin Y C, Bao M Q, Cheng R, Bai J W, Liu Y, Qu Y Q, Wang K L, Huang Y and Duan X F 2010 {\it Nature} {\bf 467} 305-8

\bibitem{Kim}
Kim S G, Luisier M, Boykin T B and Klimeck G 2011 {\it Appl. Phys. Lett.} {\bf 99} 232107

\bibitem{Logoteta}
Logoteta D, Fiori G and Iannaccone G 2014 {\it Sci. Rep.} {\bf 4} 6607

\bibitem{Zheng}
Zheng J, Wang L, Quhe R, Liu Q, Li H, Yu D, Mei W N, Shi J, Gao Z and Lu J 2013 {\it Sci. Rep.} {\bf 3} 1314

\bibitem{Chanana}
Chanana A, Sengupta A and Mahapatra S 2014 {\it J. Appl. Phys.} {\bf 115} 034501

\bibitem{Fiori}
Fiori G, Bonaccorso F, Iannaccone G, Palacios T, Neumaier D, Seabaugh A, Banerjee S K and Colombo L 2014 {\it Nat. Nanotechnol.} {\bf 9} 768-79

\bibitem{Xia}
Xia F N, Perebeinos V, Lin Y M, Wu Y Q and Avouris P 2011 {\it Nat. Nanotechnol.} {\bf 6} 179-84

\bibitem{Benali2013}
Benali A, Traversa F L, Albareda G, Aghoutane M and Oriols X 2013 {\it Appl. Phys. Lett.} {\bf102} 173506

\bibitem{bitlles}
http://europe.uab.es/bitlles/

\bibitem{Pellegrini}
Pellegrini B 1986 {\it Phys. Rev. B} {\bf 34} 5921

\bibitem{Shockley}
Shockley W 1938 {\it J. Appl. Phys.} {\bf 9} 635

\bibitem{Ramo}
Ramo S 1939 {\it Proc. of the IRE} {\bf 27} 584-5

\bibitem{Benali2012}
Benali A, Traversa F L, Albareda G, Alarc\'{o}n, Aghoutane M and Oriols X 2012 {\it Fluc. Noise Lett.} {\bf 11} 1241002

\bibitem{Novoselov2006}
Katsnelson M I, Novoselov S K and Geim A K 2006 {\it Nat. Phys.} {\bf 2} 620-5

\bibitem{Jena}
Jena D, Fang T, Zhang Q and Xing H L  2008 {\it Appl. Phys. Lett.} {\bf 93} 112106

\bibitem{David}
David J K, Register L F and Banerjee S K 2012 {\it IEEE Trans. Electron. Devices} {\bf 59} 976-82

\bibitem{Paussa}
Paussa A, Fiori G, Palestri P, Geromel M, Esseni D, Iannaccone G and Selmi L 2014 {\it IEEE Trans. Electron. Devices} {\bf 61} 1567-74

\bibitem{Zhang}
Zhang L M and Fogler M M 2008 {\it Phys. Rev. Lett. } {\bf 100} 116804

\bibitem{Oriols2001}
Oriols X, Mart\'{i}n F and Su\~{n}\'{e} 2001 {\it Appl. Phys. Lett.} {\bf 79} 1703

\bibitem{Marian}
Marian D, Colom\'{e}s E, Zhan Z and Oriols X 2015 {\it J. Comput. Electron.} {\bf 14} 114-28

\bibitem{Oriols2007}
Oriols X 2007 {\it Phys. Rev. Lett.} {\bf 98} 066803

\bibitem{Oriols2009}
 Alarc\'{o}n A and Oriols X 2009 {\it J. Stat. Mech. Theor. Exp.} {\bf 2009} 01051
 
 \bibitem{Albareda}
Albareda G, Traversa F L, Benali A and Oriols X 2012 {\it Fluc. Noise Lett.} {\bf 11} 1242008
 
 \bibitem{Oriols2013}
 Alarc\'{o}n A Yaro S, Cartoix\`{a} X and Oriols X 2013 {\it J. Phys. Condens. Matter} {\bf 25} 325601

\bibitem{Howes}
Howes M J and Morgan D V 1980 {\it Microwave Solid State Devices and Applications } (England: Peter Peregrinus Ltd)

\bibitem{image1}
Barrera R G, Guzm\'{a}n O and Balaguer B 1978 {\it Am. J. Phys.} {\bf 46} 1172-9

\bibitem{image2}
Ianovici M and Morf J J 1977  {\it IEEE Trans. on El. Ins.}  {\bf EI-12}  165 - 170


\end{thebibliography}
\end{document}